# Infiltrating a thin or single layer opal with an atomic vapour: sub-Doppler signals and crystal optics

Elias Moufarej[1,2], Isabelle Maurin[1,2], Ilya Zabkov[3], Athanasios Laliotis[1,2], Philippe Ballin[1,2], Vasily Klimov[4], Daniel Bloch[2,1,(a)]

[1] *Laboratoire de Physique des Lasers, Université Paris13- Sorbonne-Paris-Cité, 99 Av. JB Clément, 93430 Villetaneuse, France*
[2] *UMR 7538 du CNRS, 99 Av. JB Clément, 93430 Villetaneuse, France*
[3] *Moscow Institute of Physics and Technology (State University), Institutskiy Pereulok 9, 141700 Dolgoprudny, Russia.*
[4] *P.N. Lebedev Physics Institute, 53 Leninskii prospekt, 119991 Moscow, Russia*

Abstract – Artificial thin glass opals can be infiltrated with a resonant alkali-metal vapour, providing novel types of hybrid systems. The reflection at the interface between the substrate and the opal yields a resonant signal, which exhibits sub-Doppler structures in linear spectroscopy for a range of oblique incidences. This result is suspected to originate in an effect of the three-dimensional confinement of the vapour in the opal interstices. It is here extended to a situation where the opal is limited to a few or even a single layer opal film, which is a kind of bidimensional grating. We have developed a flexible one-dimensional layered optical model, well suited for a Langmuir-Blodgett opal. Once extended to the case of a resonant infiltration, the model reproduces quick variations of the lineshape with incidence angle or polarization. Alternately, for an opal limited to a single layer of identical spheres, a three-dimensional numerical calculation was developed. It predicts crystalline anisotropy, which is demonstrated through diffraction on an empty opal made of a single-layer of polystyrene spheres.

**Sub-Doppler atomic spectroscopy and confinement**

   The development of meta-materials and artificial materials, at a scale comparable to or smaller than optical wavelengths, opens large prospects to manipulate optical fields with unprecedented possibilities. Only few connections have been established with Atomic Physics until now, even though atomic transitions provide universal filters or references for light frequency. It is only very recently that cold atoms can be manipulated in the vicinity of artificial materials [1]. Rather, confined atoms or molecules are usually in a thermal gas state and undergo the spectroscopic Doppler broadening when they are confined close to artificial materials such as porous materials [2, 3] and holey or photonic fibres [4, 5]. Sub-Doppler resolution was demonstrated only through *nonlinear* schemes derived from the saturated absorption technique (well-known for large volume) for a vapour undergoing a loose confinement (not better than 10 μm), either with a counter-propagating beam [5] in a holey fibre, either [3] through the nearly counter-propagating effect of the scattered light in a porous medium.

   For a one-dimensional ("1-d") confinement with nano-cells of dilute vapour, a sub-Doppler contribution appears in *linear* spectroscopy, restricted to nearly normal incidence [6]. This is because of a transient regime of absorption, which enhances the relative contribution of the slow atoms, and which can be traced back to a special type of Dicke narrowing. Looking for a three-dimensional ("3-d") extension of these observations, we eventually observed sub-Doppler structures for the reflection from a gas confined in the interstitial regions of an opal of glass nanospheres [7].

   In the present work, we extend our previous experimental investigations [7] by improving the range of experimental conditions yielding sub-Doppler structures. In particular, we reduce the opal thickness of the confining opal. Unexpectedly, we still observe narrow sub-Doppler spectra with an opal made of a single layer arrangement of nanospheres opal.

   The detailed physics of the atomic response leading to these narrow sub-Doppler line remains elusive: it combines (i) the nano-optics problem of light propagation inside the opal, (ii) the light-atom interaction in opal voids (of a complex shape) whose response is transient and nonlocal because of the atomic motion, and finally (iii) the summing-up of atom radiation onto the direction of the reflected field. To improve our understanding of the optics which governs the excitation brought to the atoms, and the summing-up of the atomic response, we have implemented two kinds of optical models. The first one takes into account the intrinsic imperfections of our typical opals (prepared by a Langmuir-Blodgett deposition) and can be extended to the case of a resonant infiltration; the other one deals with the special case of a single layer opal, and predicts a crystalline anisotropy,

---

(a) E-mail: *corresponding author, daniel.bloch@univ-paris13.fr*

that we eventually observed for a well-organized single layer opal.

### Infiltrating an opal with Caesium vapour

An opal consists of a crystalline arrangement of monodisperse micro- or nano-spheres, usually made of glass (silica) or polystyrene. Various schemes of soft chemistry allow producing a self-organized photonic crystal [8], whose local defects are numerous because of the self-organization.

Filling up the interstitial regions of such a photonic crystal with spectroscopically convenient vapours of alkali metal -Cs in our case- is not easy as the infiltration induces cluster formation in the small interstices. This problem is exemplified in fig.1, where we show successive pictures of a thick opal (1mm thickness, sphere diameter D ~ 200 nm) after the filling-up of an evacuated sealed glass cell where the opal was introduced. When filling-up the cell, the Cs vapour condenses within the interstices of the opal. The opal looks almost black, and does not even exhibit a metallic aspect. Because the interstices are numerous, and are of a small size, it is much more difficult to evaporate the dense Cs than for a usual planar interface. Evaporated atoms are easily re-adsorbed, or interact with neighbouring atoms to generate clusters, whose signature is the greenish colouring of the opal. Moreover, when heating-up the opal to evaporate the dense Cs, thermal inhomogeneities and the dynamics of evaporation and readsorption mostly result in the displacement of the coloured regions bearing the clusters. A possible explanation of the considerable trapping of the Cs atoms in the opal interstices could be that silica nanospheres are far from being perfectly spherical as revealed by electron microscopy [9] (rather, glass speres are structured as a quasi-spherical cluster of smaller spheres).

Successful operation occurred when replacing the macroscopic opal by a thin opal prepared through a Langmuir-Blodgett (LB) layer-by-layer deposition technique [10]. LB deposition allows controlling the number of layers (presently between 1 and 20) deposited on a glass window, at the expense of the crystal quality (a LB opal is usually organized only in a random hexagonal close-packed -r.h.c.p.- structure). The window is further inserted into a cell (see fig. 2), finally sealed after filling-up with Cs. Until now, the cell thickness remains macroscopic (typically 4 cm), although it would be advantageous to fabricate a cell thin enough for the opal to be just a "spacer" for the two windows [11]. This project has been delayed until now because of residual inhomogeneities in the opal thickness on the one hand, and because of technological problems for the gluing of highly parallel windows on the other hand. Practically, a sufficient overheating of the opal-covered window resulted in a stable operation for the opal, which remains with its milky appearance (i.e. it does not absorb, but scatters light in spheres of a size comparable to the wavelength).

### Sub-Doppler reflection from the opal interface

The design of the cell containing the opal infiltrated with Cs vapour does not allow using transmission spectroscopy to probe the infiltrated opal (see fig. 2) because nearly all of the active (optically resonant) volume is filled with free vapour, rather than with the opal. A detection of the scattered light may be implemented, but should view only the region of the opal, and requires a high sensitivity. Rather, our detection relies on the spectroscopic monitoring of the light reflected at the interface between the window and the opal (see fig. 2), i.e. we look for the resonant changes $\Delta R(\omega - \omega_0)$ of the reflectivity R when the irradiation frequency is tuned across the atomic vapour resonance. The spatial origin of the signal that we monitor is a problem mostly addressed in the next section. Note that the resonances for a dilute gas are considerably narrower than any wavelength-dependent behaviour associated to the photonic crystal nature of the opal. In order to relatively enhance the narrowest contributions in the spectrum, we adopt a frequency modulation (FM) technique, where a lock-in detector yields the frequency-derivative of the reflection spectrum.

Our initial experiments [7, 12] were performed with frequency-tuneable narrow linewidth laser diodes on the doublet of resonance lines of Cs ($\lambda_1 = 0.894$ µm and $\lambda_2 = 0.852$ µm) with cells containing deposited opals made of 10 or 20 layers of 1µm diameter spheres. Observations were similar for these two opals, and no significant difference is found between the Cs density in the opal region and in the gas region (Cs density $10^{13}$-$10^{14}$ at.cm$^{-3}$). Sub-Doppler spectra were observed under near normal incidence, whose width was ~ 30-40 MHz, as opposed to ~ 400 MHz full Doppler width (natural width for Cs resonance is ~5 MHz, and the collision self-broadening is comparable to this width. In spite of a considerable broadening when increasing the incidence angle θ, the major result is that narrow sub-Doppler structures, superimposed to a broad spectrum, appear for a limited range of (large) oblique incidences (θ ∼ 40-50°). The contrast is higher for TM-polarisation, and for a given incidence, the shape of this structure strongly differs for TE or TM polarizations. Also, for a given polarization, the shape and amplitude evolve quickly with the incidence angle. In addition, major differences appear when comparing neighbouring wavelengths such as $\lambda_1$ and $\lambda_2$. All these original observations were made in the pure *linear* regime of resonant interaction. Our simple interpretation was that close to normal incidence, the narrow spectra are rather analogous to the ones of a thin cell [6] because of the large voids between the window and the first half-layer of spheres, while the sub-Doppler structures observed for a range of oblique incidences were a genuine consequence of the 3-d confinement of the atomic vapour.

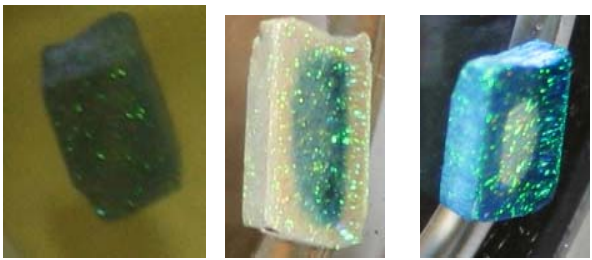

Fig. 1 : The 1mm-thick opal, totally opaque after filling-up the cell with Cs (left), becomes (centre) inhomogeneous and greenish -in the course of evaporation obtained by selective heating of the opal region- ; it returns (right) to a darker colour when the heating-up of the window is stopped, with colouring occurring first on the outer part.

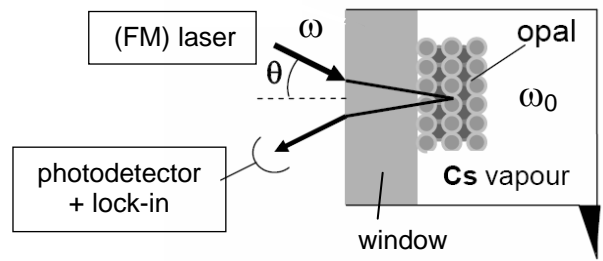

Fig. 2: Scheme of reflection spectroscopy at the interface between a Cs cell window, and a LB opal deposited on it. For convenience, the size of the thin opal (made here of 5 layers in a compact arrangement) is strongly exaggerated. The frequency of the incident beam ω is scanned across the Cs resonance $\omega_0$. θ is the incidence angle in air before the window.

We have recently conduced complementary experiments with new cells and different sphere diameters (D ∼ 400 nm). To study the weak Cs line at λ = 455 nm (see fig.3a), the sensitivity has been enhanced by compensating for a residual amplitude modulation induced by the FM laser. In addition to well-resolved hyperfine components close to normal incidence (*e.g.* 3°), we find ranges of oblique incidences (see *e.g.* 36°) where relatively narrow structures mix-up with broader features associated to a full Doppler width (∼ 800 MHz). Their shape varies quickly with the incidence angle. Polarization dependence is always strong, with TM polarization generally more favourable, perhaps because non resonant reflection is weaker in the vicinity with the Brewster angle. For larger spheres (D ∼ 1 μm) and λ = 455 nm, rather broad structures, however well below the full Doppler width, have been also observed for a limited range of incidence angles.

We have also acquired sealed Cs vapour cells bearing a multi-zone window (covered by stripes of opals made with *e.g.* 1, 2, 3 and 4 layers). With these few layer opals, narrow structures are still observed on the Cs $D_1$ line far away from normal incidence (fig. 3b). Their amplitude is relatively larger than for a many layer opal and can hinder the Doppler-broadened structure. Remarkably, in the limiting case of a single layer of glass spheres, a sub-Doppler lineshape remains observable in TM polarization. This novel and polarization-dependent result is unexpected as it cannot be traced back to a 3-d confinement.

In all these experiments, the lineshapes are highly sensitive to light polarization. For experiments as ours which are performed in the linear regime (with respect to intensity), only the complex propagation inside the opal (and not the atom excitation) can be affected by the light polarisation. This is why we have addressed the problem of the optical propagation in the opal through modelling, and performed complementary experiments on non infiltrated opals. We report below on two different types of optical approaches: one, well-suited to a LB opal of several layers, allows considering a resonant infiltration; the other one, specific to a single layer, makes the geometrical orientation an important parameter.

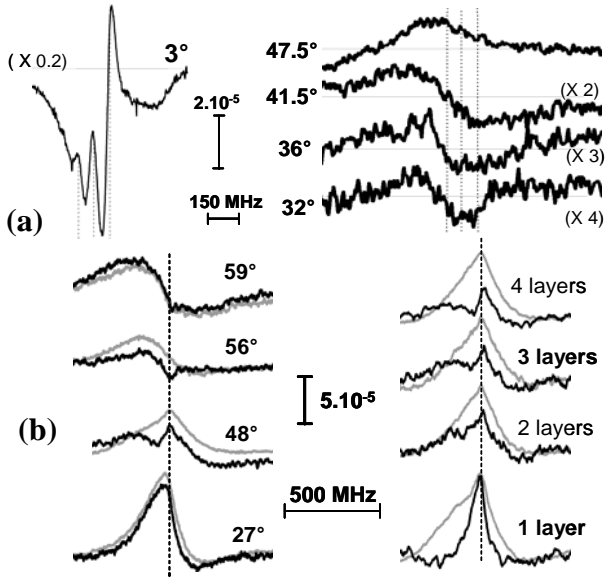

Fig. 3: FM reflection spectra on Cs at an opal interface. For all spectra, the vertical (dashed) lines mark atomic resonances (provided by auxiliary saturated absorption spectra). The vertical scale provides ΔR/R for a 16 MHz amplitude of the FM. (a) experiments on the λ = 455 nm line at T = 140 °C (∼ $1.5.10^{14}$at/cm$^3$) for an opal made of 10 layers of D = 400 nm; incidence angles as indicated, TM polarization; (b) experiments at $\lambda_1$ = 894 nm at T = 100°C (∼ $2.10^{13}$at/cm$^3$) on a multizone opal (D = 1030 nm) and TE (grey) or TM (black) polarizations. Spectra on the left are for a 3 layers zone, incidence as indicated and on the right for a 45° incidence, and number of layers as indicated.

**Layered model of the opal and resonant infiltration**

Many models were developed to understand the optics of photonic crystal. They are often based upon heavy numerical bandgap calculations, and do not take into account the many defects and partial disorder of a LB opal. They are not well-suited to include a perturbative change of parameters, such as the λ/D ratio, or the sphere or substrate index. To overcome these difficulties, we have developed [13] a one-dimensional (1-d) model, based upon a transfer matrix calculation analogous to a multi-layer thin film, with the layered effective index $n_{eff}(z)$ described by:

$$n_{eff}(z) = \{(f(z)n^2_{sphere} + [1-f(z)]\}^{1/2} \quad (1)$$

In eq. (1), z is the distance to the window, and f(z) the filling factor of the opal in the plane at height z. The structure of $n_{eff}(z)$ is shown in fig. 4 a. To simulate the scattering in the transparent opal, an additional extinction parameter is added. This flexible model easily allows *quantitative* predictions, notably of the width and amplitude of the Bragg rejection peak, and even of high-order Bragg peaks. Dedicated tests with a variety of opals deposited on a window have shown a reasonable success against experimental data, which was acquired with a white source and a spectrometer [13]. The intrinsic disorder of the *r.h.c.p.* arrangement of a LB opal, and dispersion in the glass sphere sizes justifies the success of this crude 1-d model replacing a complex 3-d photonic crystal.

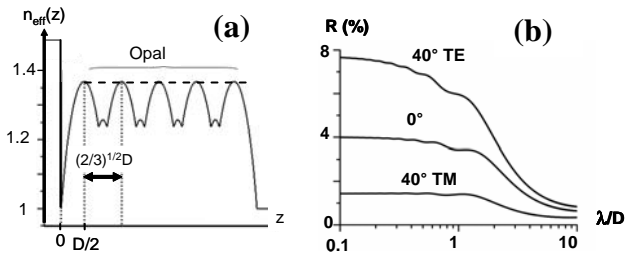
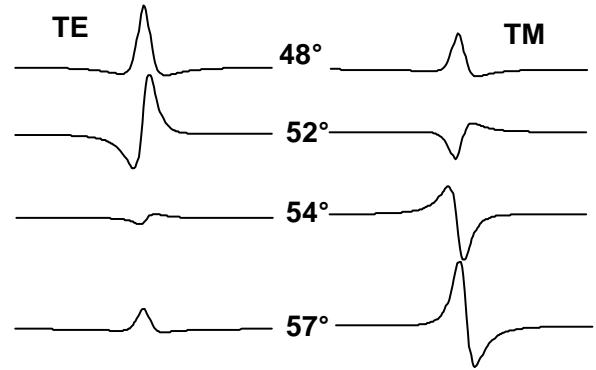

Fig. 4 : (a) Effective index model used to describe the opal as a layered medium; the horizontal dashed line is for a "fused" opal model. (b) Calculated reflection at the window/"fused" opal interface as a function of $\lambda/D$ for normal incidence or for $\theta = 40°$ and TE or TM polarization as indicated. Index for the glass nanospheres is 1.4, and 1.5 for the ($z < 0$) window.

This model also helps to interpret the physics underlying the predicted behaviours. In particular, it includes quite naturally the break of the compact arrangement for the first and last half-layers. This exemplifies a very general problem encountered with photonic crystals: optical measurements require a coupling to the outside regions – *e.g.* by transmission or reflection-, which breaks the periodicity (on a $\lambda$ scale) at the interface. In fig. 4b, a calculation is performed for a "fused" opal in order to enhance the "gap" effect between the substrate and the first -layer. It shows that the reflectivity from the interface is very close to the one for a substrate/vacuum interface when the irradiation wavelength remains comparable to the opal size (*i.e.* for $\lambda/D \leq 1$, reflectivity is the same as for $\lambda/D \rightarrow 0$, which is the prediction for an elementary planar interface). This is in agreement with experiments [7, 13].

This model, based upon a linear matrix formalism, can be extended [14] to describe an opal infiltrated by a low density resonant medium (*e.g.* a gas of motionless atoms or a liquid containing a resonant doping). For this purpose, one simply assumes the resonant infiltrated medium to be characterized by an index $n_{infilt} = 1 + \delta n$, and extrapolates the layered model to a first order expansion in $\delta n$. The density of the resonant dilute medium is assumed to evolve periodically, following the density of voids in the opal [1-f(z)]. Calculating the effect of an infinitesimal slice of resonant medium, one finds that the resonant change in the reflectivity oscillates with the spatial position of the slice. The period of this oscillation depends on the wavelength and on the incidence angle. Its phase differs for the contributions originating in Re($\delta n$) or in Im($\delta n$) (respectively dispersive and absorptive contribution); it also varies with light polarisation (for large incidences). Hence, when integrating the elementary responses over the distribution of the infiltrated material, it is only for specific incidence angles that this spatially-oscillating response matches the periodical emptiness factor [1-f(z)], governed by the opal geometry. In this case, a large resonant signal is predicted, owing to a constructive contribution of the response of the whole infiltrated material. This shows that the origin of the resonant reflection extends over several wavelengths, *i.e.* over the deep layers of the opal. For a resonant infiltrated material characterized by a complex dependence

Fig 5: Calculated FM frequency lineshapes for $\Delta R/R$ around resonance for an infiltrated opal. The complex index $\delta n$ of the infiltrated material is assumed to be governed by a Lorentzian resonance. Calculation for a 20 layers opal, $\lambda/D=0.852$, and a scattering analogous to an absorption 0.2 $D^{-1}$. Polarization and incidence angles as indicated, all curves represented with the same scale, For 48°, 52°, 54°, 57°, and respectively (TE, TM) polarizations, one finds R = (5.9 %, 0.45 %), (10.3 %, 0.29 %), (7. 3%, 0.14 %), (5.9 %, 0.08 %).

Re[$\delta n(\omega-\omega_0)$] and Im[$\delta n(\omega-\omega_0)$], it is the overall lineshape of the resonant reflectivity which varies quickly with the incidence angle and with the incident polarization. This is observed in fig.5, where a complex Lorentzian resonance is assumed. Such a modelling provides a simple analogy with the quick variations, experimentally observed in a gas, with the incidence angle and polarization. It is however too limited to deal with the complexity of a transient velocity-dependent atomic resonance, and to justify sub-Doppler widths.

Figure 5 shows relative resonant changes $\Delta R/R$, because it allows an easier comparison between TE and TM polarizations. Despite the limited range of incidence angles, the variations of the non resonant reflectivity R for a given polarization are non negligible in fig.5. This is because of a simultaneous second-order Bragg peak of reflectivity. This coincidence, surprising only at first sight, is due to the fact that the condition for a matched contribution of the resonant slices is analogous to the conditions yielding a Bragg-type resonance [14]. This does not mean that the resonant variations which are predicted are just an effect of this (weak) photonic crystal mode. Indeed, the lineshapes predicted in fig.5 are highly sensitive to the assumptions chosen to evaluate the resonant response. In particular, we have ignored the optical excitation induced by the resonant scattered light, which should affect the phase of excitation in the infiltrated medium.

**Single layer opal and crystal optics**

The layered effective index model is not well-suited for a single or double layer opal. Conversely, a rigorous numerical calculation -by finite elements-, taking into account the periodicity associated to a perfect crystal, has been developed by I. Zabkov *et al.* (unpublished). By truly taking into consideration the shape and periodicity of the contacting spheres, the calculation should provide a quantitative evaluation of the reflection and transmission coefficients, and of the scattering as well, with no need for an *ad hoc* extinction coefficient. In this nano-optics approach, the crystal orientation φ, defined

relatively to the plane of incidence (see fig. 6), is a key parameter of the calculation

The model predicts rapid variations of reflectivity with the opal orientation for some wavelengths (see fig.6). However, general trends (*e.g.* a minimum around 530 nm and 700 nm) are easily found. They are probably associated to a kind of resonance between the wavelength and the optical size of the sphere. We have performed experiments on a single layer of glass spheres (D=0.73μm) deposited on a microscope slide, using fibered "white laser" source and a spectrometer, as described in [13]. Experimentally, opals made of glass spheres are highly poly-domain because of dispersion in sphere diameter. This implies to include in the model a numerical averaging over the orientation angles. Very good agreement with experiments is hence found with this averaging, for reflection and transmission as well (fig.6). Note that transmission appears less sensitive to the opal orientation.

Until now, our opals made of glass spheres through a LB technique have not been able to exhibit the diffractive behaviour [11,15] associated to the hexagonal distribution of spheres in an elementary opal layer. This is because of a high dispersion in sphere diameter (up to ~ 5 %) for glass spheres. Conversely, for polystyrene spheres, single domains are found on large areas owing to a much smaller dispersion (~ $10^{-3}$). The hexagonal diffraction structure, usually turned into a circle for a highly poly-domain opal, has been easily observed with visible light when focusing our blue laser at 455 nm onto an opal with D = 0.28 μm spheres. High-order diffraction spots could even be observed. Under normal incidence, there is no TE or TM polarization relatively to the incidence plane, but a preferential polarization direction can be expected relatively to the opal orientation. In other words, the intrinsic anisotropy of the opal makes it optically a crystal. As shown in Fig. 7, the diffraction efficiency for λ = 455 nm varies by ~ 30% under normal incidence when changing polarization. Smaller variations (~10 %) are found for a longer wavelength (green He-Ne laser at λ=543 nm).

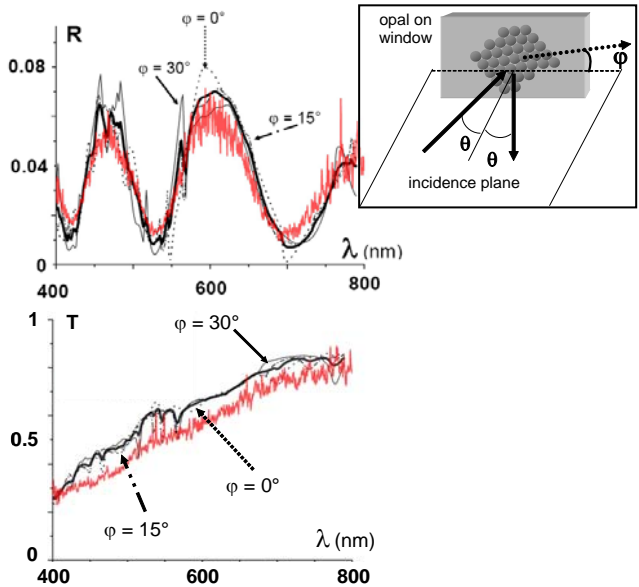

Fig.6: Reflection (R) and transmission (T) spectra of a single layer of glass opal. The inset defines the irradiation incidence angle (θ) and the opal orientation (φ) angle. The experiment (red, noisy) is performed with D = 0.73 μm glass spheres, θ = 15°, polarization is TM. The calculations (black) are for a dimensionless parameter λ/D and adjusted to the effective D value. Various orientations of the opal are considered (as indicated), and the thick black line is for a randomized orientation.

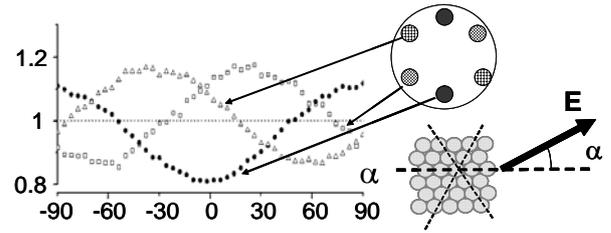

FIG.7: Relative intensity of a pair of diffracted spots as a function of the orientation of the linear polarisation for an irradiation inder normal incidence. The direction of polarisation α (in degrees) is defined by the opal orientation (see bottom right). This relative intensity is normalized by the average hexagonal diffraction efficiency. The circular inset (top right) shows different pairs of equivalent spots for an opal oriented as shown. The experiment is performed on a single layer of polystyrene spheres (D = 0.28 μm, λ = 455 nm).

The numerical model has the ability to predict the diffraction efficiency and its anisotropy. A comparison with our results would provide a sensitive test of the presence of opal defects, affecting the mesoscopic organization of the spheres, or the microscopic geometry of individual spheres [9].

As previously discussed, *transmission*, used here to monitor diffraction, cannot be presently applied to an opal infiltrated with a vapour. Meanwhile, we have also observed, hexagonal diffraction *in reflection* on a single layer polystyrene opal, although with a weaker efficiency. Hence, extension to a single-layer infiltrated opal essentially requires the feasible preparation of a single domain *glass* opal [11].

## Conclusion

Artificial glass opals prepared in thin layers can be used to confine atomic vapours. They offer a simple example of a hybrid system interfacing atoms with artificially structured materials. Sub-Doppler spectroscopic signals are found for a gas infiltrated opal. A generalization of these results has even been found for a single layer opal, which can be viewed as a kind of 2-dimensional grating deposited on the window. A major step for the full understanding of these signals would be to obtain a detailed description of the optical propagation inside the opal, which is usually a partially disordered photonic crystal. A simplified layered model has allowed understanding the strong influence of the gap region between the substrate and the contact region of an opal or more generally of a photonic crystal, and the possibility to observe the response of an in-depth infiltration. A detailed three-dimensional treatment for a single layer opal has revealed crystal optics properties, experimentally observed through polarization-sensitive diffraction. This may allow versatile detection schemes based upon the simultaneous detection of diffracted and reflected beams, opening prospects of new insights to analyze the origin of the sub-Doppler contributions in infiltrated opals.

**Acknowledgements**

Work supported by ANR [08-BLAN-0031 "Mesoscopic gas"], and partly by CNRS-PICS 5813. We thank A. Maître INSP-Paris) for providing us with the thick opal, S. Ravaine (CNRS Bordeaux) for the LB opals, and F. Thibout (LKB-Paris) for fabricating the vapour cells.